\documentclass{article}
\usepackage{authblk}
\usepackage{subcaption} 
\usepackage{gnuplottex} 
\usepackage{epstopdf} 
\usepackage{graphicx}
\usepackage{amsmath}
\usepackage{graphicx}
\usepackage{balance}
\usepackage{comment}

\usepackage{booktabs} 
\usepackage{epsfig}
\usepackage{url}
\usepackage{amsmath}
\usepackage{epstopdf}
\usepackage{enumitem}
\usepackage{algorithm}
\usepackage{todonotes}

\usepackage{subcaption} 
\usepackage{gnuplottex} 
\usepackage{epstopdf} 
\usepackage{graphicx}
\usepackage{latexsym}
\usepackage{keyval}
\usepackage{ifthen}
\usepackage[utf8]{inputenc}

\title{An Anonymous Overlay Routing Protocol for Uplink-Intensive Applications}
\author[1]{Francesco Buccafurri}
\author[1]{Vincenzo De Angelis}
\author[1]{Sara Lazzaro}
\affil[1]{University of Reggio Calabria, Via dell'Università 25, Reggio Calabria 89124, Italy}
\affil[ ]{\textit {\{bucca,vincenzo.deangelis,sara.lazzaro\}@unirc.it}}
\date{}                     
\begin{document}

\maketitle
\begin{abstract}
Sender anonymity in network communication is an important problem, widely addressed in the literature.
Mixnets, combined with onion routing, represent certainly the most concrete and effective approach achieving the above goal. 
In general, the drawback of these approaches is that anonymity has a price in terms of traffic overhead and latency.
On the Internet, to achieve scalability and not to require
relevant infrastructure and network-protocol changes, 
only P2P overlay protocols can be adopted.
Among these, the most representative proposal is certainly
Tarzan, which is designed to obtain strong anonymity
still preserving low-latency applications.
In recent years, we are witnessing a change in Internet traffic.
Due to IoT, cloud storage, WSN, M2M, uplink traffic is more and more increasing. An interesting question is whether this new traffic configuration may enable
new strategies to improve the effectiveness of Tarzan-like approaches.
In this paper, we investigate this problem, by proposing \textit{C-Tarzan}, an anonymous overlay P2P routing protocol.
Through a deep experimental analysis, we show that C-Tarzan
outperforms Tarzan in the case of
uplink-intensive applications.
\end{abstract}

\section{Introduction}
\label{sec:introduction}


Anonymity in network communication is a widely investigated problem \cite{shirazi2018survey}.
Obviously, it is not sufficient to hide the content of exchanged messages, 
as data related to traffic carry out sensitive information per se. Thus, anonymous communication networks (ACNs) aim to offer a certain degree of 
unobservability of communication in the network,
not just message confidentiality.
The most known and used anonymous protocol is Tor \cite{dingledine2004tor}. However, as well-known, anonymity is easily broken under even weak
threat models \cite{karunanayake2020anonymity}.
A challenging goal is to guarantee robust
sender anonymity because it is enough to achieve relationship anonymity
\cite{pfitzmann2010terminology}.
For robust, we mean that
both passive sniffers and malicious participants cannot 
distinguish whether a node generates a message or simply relays it.

The most effective approaches existing in the literature achieving the above goal are based on the concept of mixnet \cite{chaum1981untraceable} including cover traffic.
Mixnet protocols rely on intermediate servers (called \textit{mix-nodes}) that mix the messages coming from different sources to hide the relationship between the incoming messages to and the outcoming messages from the mix-nodes.
When cover traffic is included in mixnets, serious problems of
traffic overhead may arise.
While a wide literature regarding mixnets exists, a few proposals
mixnet-based oriented to a concrete Internet (low-latency) implementation of the notion of mixnet, including cover traffic, are available.
Among these, if we refer to P2P approaches
(thus not requiring infrastructure changes), the most meaningful proposal
is certainly Tarzan \cite{freedman2002tarzan}.

Despite its age, Tarzan is the only effective proposed anonymous routing protocol guaranteeing low latency even in large scale Internet scenarios. Indeed, the protocol allows a client to anonymously contact a server through a tunnel whose length is independent of the number of nodes participating in the peer-to-peer network. As a matter of fact, Tarzan implements a peer-to-peer overlay network at IP layer, in which peers collaborate with each other to implement anonymous tunnels through which a client may reach a proxy node (called PNAT) from which the server is reached.
Another advantage of Tarzan with respect to recent state-of-the-art approaches is that, unlike the emerging mixnets that adopt centralized and explicit shuffling nodes (\cite{piotrowska2017loopix}),
the peer-to-peer approach makes the solution more robust against possible attacks on the nodes of the route (or their collusion). 
Indeed, all the nodes of the network are potentially sender or relay nodes and then there are no few explicit targets for the attacker.
The most recent (and representative) approach using a peer-to-peer overlay network is \cite{DAENET}. However, \cite{DAENET} does not work at IP layer and, moreover, the length of each communication path is $log \ n$, where $n$ is the number of nodes of the network. Therefore, unlike Tarzan, the latency is growing with the number of nodes.
Hence, the protocol is not suitable for low-latency applications when the number of users scales at huge values, as may happen in Internet scenarios.

The aim of this paper is to understand whether the change of
type of Internet traffic due to various reasons (emerging applications for IoT, M2M, cloud, etc.), for which uplink traffic is more and more increasing \cite{oueis2016uplink,yang2019can,shafiq2013large,berger2015joint}, might allow us to find
some improvement to the Tarzan approach to make it
more suitable to the new scenario.

The study conducted in this paper leads to the definition of a new
P2P overlay anonymous protocol, called  \textit{C(yclic)-Tarzan},
which outperforms Tarzan in the case of uplink-intensive applications.
The core idea is that the topology of the overlay network
allows us to set in the network just unidirectional cover traffic
instead of the bidirectional traffic required in Tarzan.

Our study is based on
the well-known trilemma, called the \textit{anonymity trilemma} \cite{das2018anonymity}, which states the existence of a trade-off between three metrics: the anonymity set size, the latency, and the cover traffic level.
Specifically, we show that for uplink-intensive applications, by fixing the same latency and the same cover traffic volume, C-Tarzan offers a greater anonymity set size than Tarzan.

The paper is organized as follows.
In Section \ref{sec:related}, we investigate the related literature. In Section \ref{sec:background}, we provide the background notions about the Tarzan protocol. Next, in Section \ref{sec:problem_formulation}, we compare Tarzan with a sketched idea of our solution, by highlighting what motivated us to investigate in this direction. The detailed protocol is presented in Section \ref{sec:C-Tarzan}. We perform an analytical study of the latency in Tarzan and C-Tarzan in Section \ref{sec:latency} and provide an experimental validation of our approach in Section \ref{sec:experiments}. Finally, in Section \ref{sec:conclusion}, we draw our conclusion.

\section{Related Work}  \label{sec:related}
Anonymous Communication Networks (ACN) \cite{xia2020balancing,shirazi2018survey} are networks in which users are provided with anonymity services protecting their privacy.
An ambitious goal to achieve is to offer anonymity guarantees against 
passive eavesdroppers (including a global adversary) and malicious participants. 
As stated in \cite{danezis2008survey}, to achieve this goal, dummy traffic needs to be injected into the network to hide the actual traffic.

In the literature, three main approaches leveraging dummy traffic
are available.
The first is based on \textit{buses} \cite{hirt2008taxis,beimel2003buses,young2014drunk}.
In this solution, a predetermined route is used by the sender to anonymously communicate with the destination. However, this technique is not scalable on a large network, since it requires an Eulerian path passing through all the nodes that leads to a prohibitive cost in terms of latency.

A second approach is represented by \textit{DC-Nets} \cite{chaum1988dining}, which offer cryptographic guarantees of anonymity, but they suffer from scalability problems as buses \cite{shirazi2018survey}.

The third approach is represented by the \textit{mixnets} \cite{chaum1981untraceable,kotzanikolaou2017broadcast,guirat2021mixim} which, in general, offers a lower latency with a price in terms of cover traffic.

Some recent mixnet proposals exist \cite{kotzanikolaou2017broadcast,van2015vuvuzela,piotrowska2017loopix,10.1145/3132747.3132783}. 
Anyway, some drawbacks should be taken into account.
For example, as recently stated in \cite{alexopoulos2017mcmix}, the work proposed in \cite{kotzanikolaou2017broadcast} suffers from very large communication overhead.
Regarding \cite{van2015vuvuzela}, as stated by the authors themselves, the end-to-end latency is about 37 seconds, which may result too high for several applications.
Moreover, these approaches rely on a server-oriented architecture, which is known to be 
less robust against possible attacks on the nodes of the route (or their collusion) \cite{DAENET} and
less scalable than P2P architecture \cite{shirazi2018survey}.

Therefore, the state of the art of P2P approach for low-latency applications is represented by Tarzan \cite{freedman2002tarzan}, which is a work with high impact in the (even current) scientific literature.

Actually, another P2P mixnet proposal, less recent but adopted in practice, is I2P \cite{zantout2011i2p}. However, it suffers from different vulnerabilities such as brute-force attacks or timing attacks. Then, as reported in the official website ({\tt geti2p.net}), the authors suggest to adopt some mitigations (e.g., constant-rate cover traffic) present in \cite{freedman2002tarzan}.

Our paper strongly refers to \cite{freedman2002tarzan}, by proposing an extension improving Tarzan in the case of uplink-intensive applications.
To the best of our knowledge, no proposal outperforming Tarzan is available in the state of the art.
On the other hand, the considered domain is
relevant. Indeed, uplink-intensive applications are becoming more and more
common in recent years  \cite{oueis2016uplink,yang2019can}.
Some examples of uplink-dominant applications are represented by M2M \cite{nikaein2014openairinterface,centenaro2015study}, Industrial IoT \cite{kwon2016dominant}, and Wireless-Sensor-Network \cite{dester2018performance}.
Furthermore, intrinsically, cloud-based applications increase the uplink bandwidth demand with respect to traditional client-server applications \cite{sun2020adaptive}.

\section{Background: The Tarzan Protocol} \label{sec:background}
In this section, we provide the technical background about the Tarzan Protocol \cite{freedman2002tarzan}. We focus just on the main aspects useful to understand the approach we propose in this paper.

Tarzan is a peer-to-peer anonymous IP network overlay. It offers a  degree of anonymity against both a number of malicious nodes and a global adversary able to observe the entire traffic exchanged in the network.

Each node, in order to communicate anonymously with a destination, builds a tunnel composed of a sequence of nodes in which the last node communicates with a special node, called \textit{PNAT}, which acts as a proxy towards the destination.

Each intermediate node of the tunnel acts as a relay by forwarding the messages coming from the previous node. Anyway, since it does not know its position in the tunnel, it is not able to identify the originator of the traffic.

In Tarzan, the construction of the tunnel (i.e., the choice of the intermediates nodes) is not left entirely free to the initiator, which has to satisfy some constraints.

Specifically, each node is associated with a group of nodes called \textit{mimics}. To build the tunnel, the initiator $a$ chooses as first relay one of its mimics, say $b_i$. Then, $b_i$ communicates to the initiator the set of its mimics, and $a$ will choose the second relay of the tunnel among the nodes of this set. This procedure is iterated until the tunnel reaches a certain length.

To send messages through this tunnel, the initiator needs to exchange a symmetric key with each node of the tunnel. This procedure is similar to the construction of a virtual circuit in the Tor Protocol \cite{dingledine2004tor}.
To do this, the initiator first exchanges a symmetric key with the first relay of the tunnel, then it exchanges a symmetric key with the second relay through the first relay, then it exchanges a symmetric key with the third relay through the first two relays of the tunnel built so far, and so on. To exchange a symmetric key with a relay, the initiator encrypts it by using the public key of such a relay.
In such a way, no node of the tunnel can tell with whom it is exchanging the key.

Once exchanged these keys, the messages can be sent through the tunnel encrypted in a layered fashion. This means that the initiator first encrypts the message with the key of the last relay of the tunnel, then it encrypts the result with the key of the second-last relay, and so on. A relay receiving a message removes its layer of encryption and forwards the message to the next relay.

A key role in the Tarzan Protocol is played by the selection of mimics. Tarzan relies on a gossip protocol, so that each node can discover the other peers of the networks. Among these peers, the node has to select $k$ mimics. Observe that, since each node selects $k$ mimics, we can expect, on average, that it is selected as a mimic from other $k$ nodes. Therefore, the average number of mimics for each node is $2k$.

A node establishes, with each of its mimics, a bidirectional cover traffic flow into which real data can be inserted, indistinguishable, from dummy traffic. To do this, a symmetric hop-by-hop key is exchanged when a node connects to a new mimic and all the traffic exchanged between these two nodes will be encrypted with such a key.

The bidirectional cover traffic guarantees the anonymity of the senders against the global adversary and traffic analysis attacks.
Moreover, the bidirectional flow of traffic allows us to use, for the response, the same tunnel utilized to forward the request. In this case, each node crossed by the response adds a layer of encryption to each message by using a symmetric key shared with the initiator. When the initiator receives the response, it removes all the layers of the encryption.

In Tarzan, each node maintains a three-level hierarchy dynamic hash table (DHT) in which the nodes are inserted in a given position according to their IP addresses.

This table offers a $lookup$ function that, given a string as input, returns as output an IP address of a node of the network.
Observe that the input can be any arbitrary string. 

To select $k$ mimics, each node $a$ invokes the function \\$lookup^i(a.ipaddr)$ for $1 < i \leq k+1$ where $a.ipaddr$ represents the IP address of $a$.

The DHT offers two advantages. First, since the DHT is shared by all the nodes, mimic selection is publicly verifiable and then this prevents an adversary node from selecting more than $k$ mimics. 
The second advantage is that the mimics for a node are randomly selected in different IP domains, so that if an adversary controls an entire domain by generating a huge number of malicious nodes in that domain, it does not increase the probability that a malicious node of such domain is selected as a mimic.

To conclude this section, we discuss the anonymity degree achieved in Tarzan against a malicious node in the tunnel. This degree can be measured in terms of \textit{anonymity set} that is the set of potential initiators of the traffic. The anonymity set size, besides depending on the number of mimics (i.e., \textit{degree}) of the node, increases exponentially with the length of the tunnel.

\section{Problem Formulation and Basic Approach} \label{sec:problem_formulation} 

In recent years, we observed an increase in uplink traffic demand \cite{oueis2016uplink}. A lot of uplink-intensive applications emerged in different fields such as cloud-enabled ecosystem \cite{sun2020adaptive}, IoT networks \cite{kwon2016dominant}, sensor and actuator networks \cite{dester2018performance}, and so on.

In this paper, we address the problem of guaranteeing a measurable degree of anonymity in uplink-intensive applications. 
A solution could be to apply the Tarzan protocol, discussed in the previous section, in which cover traffic is adopted to offer anonymity guarantees against a global adversary.
On the other hand, the cover traffic represents overhead which results in a waste of bandwidth and energy consumption.

In Tarzan, there are three main metrics to consider \cite{das2018anonymity}: latency, the amount of cover traffic, and the size of the anonymity set. Often, the latency is a project constraint as well as the anonymity degree. 
Therefore, to find a solution that, under the same cover traffic level (that cannot be increased for the above reasons) and a fixed latency, offers a better anonymity degree than Tarzan represents an advancement of the state of the art.

Roughly speaking, we consider, as a measure for the cover traffic, the degree of the nodes.
Indeed, the more links occur in the network the more cover traffic has to be generated. Moreover, Tarzan requires bidirectional cover traffic in each link, otherwise a significant reduction of the anonymity set arises \cite{freedman2002tarzan}.

Therefore, a challenge could be to eliminate the bidirectionality of cover traffic still preserving the Tarzan-like approach.
This is the purpose of our proposal.
The idea is that unidirectional traffic could still be enabled in Tarzan protocol by rearranging node mimics in such a way that they form a cycle.
Once mimics are so organized, we can build a tunnel as in Tarzan, but requiring that
two adjacent nodes in the tunnel belong to a cycle. This way, the response can be routed by
moving back, at each hop between two nodes, by travelling the entire cycle involving these nodes. Thus, no bidirectional traffic is needed.

\begin{figure}[htp]
    \centering
    \includegraphics[width=8cm]{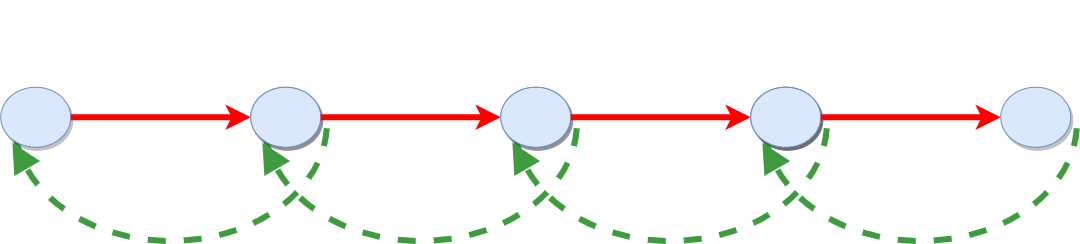}
    \caption{Forward path (red arrow) and return path (green arrow)}
    \label{fig:idea}
\end{figure}

This idea is sketched in Figure \ref{fig:idea}, in which the red lines represent the forward path and the green lines represent the cycles travelled by the response.

However, there might be a price in terms
of latency to pay when applying this cyclic approach, since, in general, the response would go through a longer path than the forward path.
Instead, in Tarzan, forward and return paths are the same.
Therefore, a solution based on the above idea is not trivially
applicable.

The first immediate consideration is that
it is convenient to minimize the size of cycles.
Being Tarzan bidirectional links 
equivalent to 2-nodes cycles, the minimum dimension for
non-trivial cycles is the case of 3-nodes cycles.
On the other hand, it is intuitive to understand that
no advantage can derive from having bigger cycles.

A much less clear point is to understand whether
we have to pay a price also in terms of anonymity set.

This question derives from the following qualitative
analysis.

\begin{figure}
     \centering
     \begin{subfigure}[b]{0.2\textwidth}
         \centering
         \includegraphics[width=\textwidth]{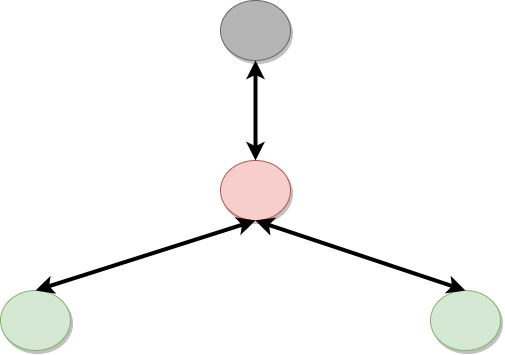}
         \caption{Tarzan topology with in-degree=out-degree=3}
         \label{fig:Uncertainty2hops_tarzan_degree3}
     \end{subfigure}
     \hfill
     \begin{subfigure}[b]{0.2\textwidth}
         \centering
         \includegraphics[width=\textwidth]{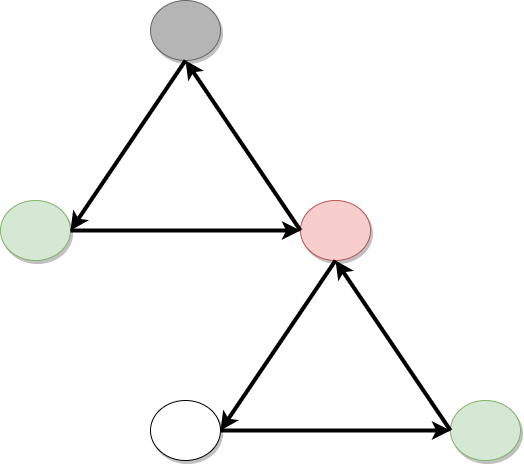}
         \caption{Cyclic topology with in-degree=out-degree=2}
         \label{fig:Uncertainty2hops_ctarzan_degree2}
     \end{subfigure}
     \hfill
     \begin{subfigure}[b]{0.2\textwidth}
         \centering
         \includegraphics[width=\textwidth]{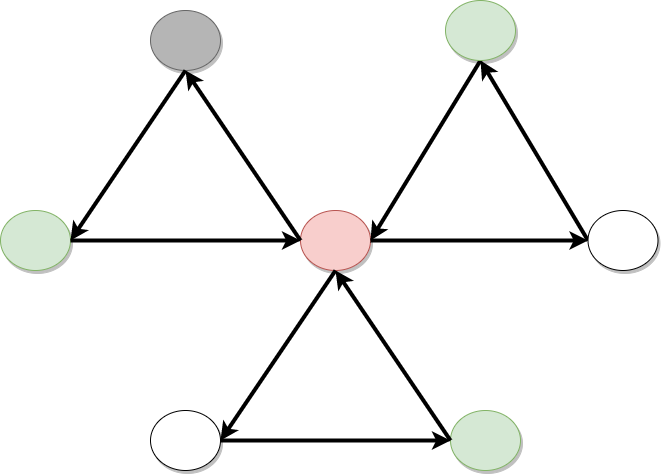}
         \caption{Cyclic topology with in-degree=out-degree=3}
         \label{fig:Uncertainty2hops_ctarzan_degree3}
     \end{subfigure}
        \caption{Uncertainty at two hops}
        \label{fig:Uncertainty2hops}
\end{figure}

\begin{figure}[htp]
    \centering
    \includegraphics[width=8cm]{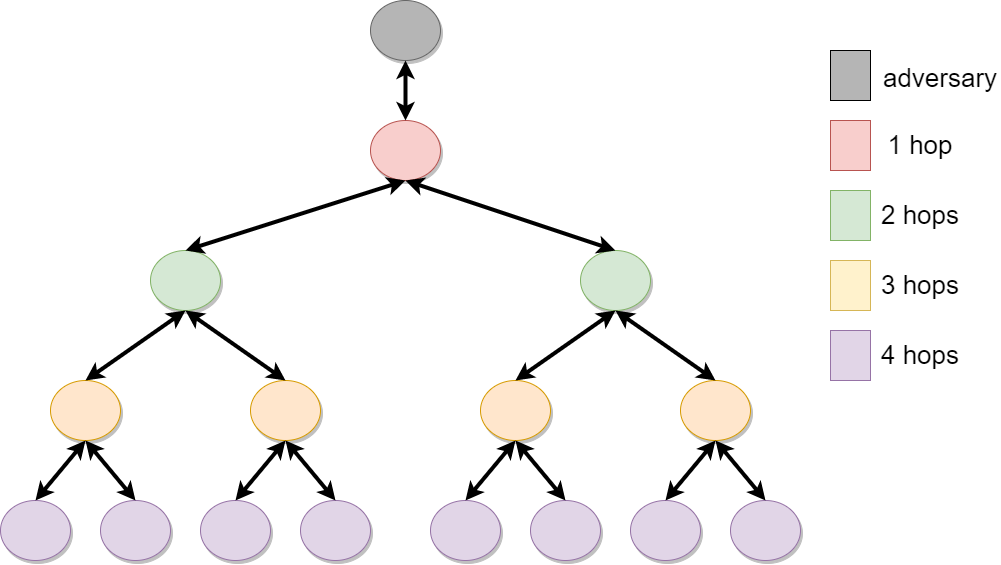}
    \caption{Extension of Figure \ref{fig:Uncertainty2hops_tarzan_degree3}}
    \label{fig:TarzanAS}
\end{figure}

\begin{figure*}[htp]
    \centering
    \includegraphics[width=14cm]{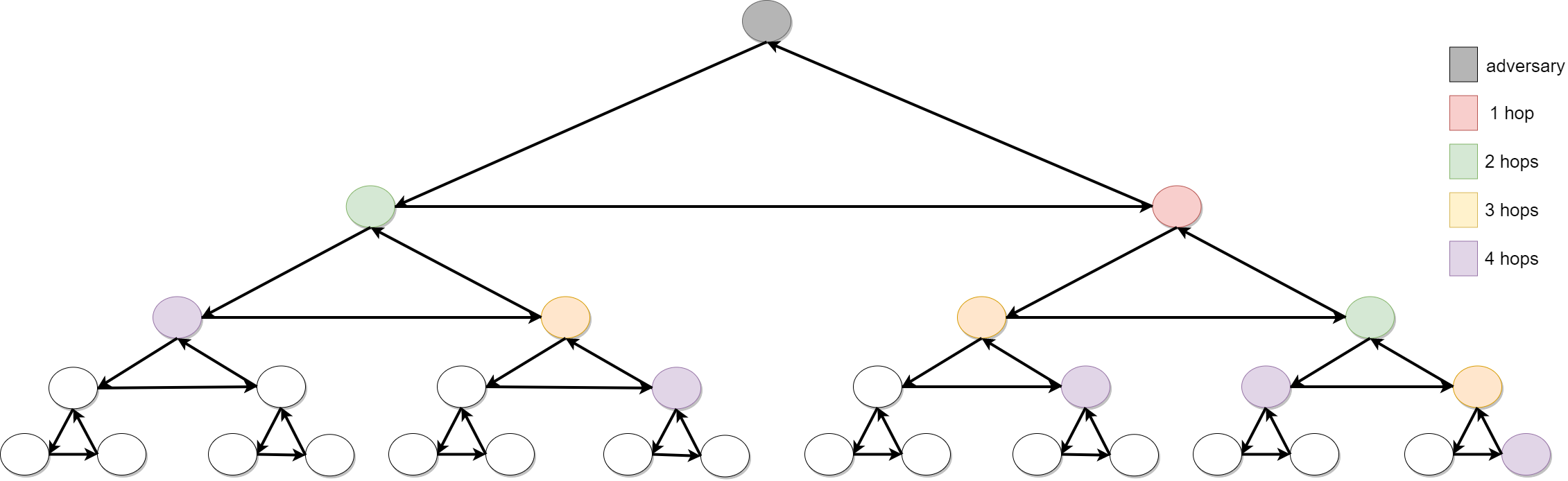}
    \caption{Extension of Figure \ref{fig:Uncertainty2hops_ctarzan_degree2}}
    \label{fig:CTarzanAS_2}
\end{figure*}

\begin{figure*}[htp]
    \centering
    \includegraphics[width=14cm]{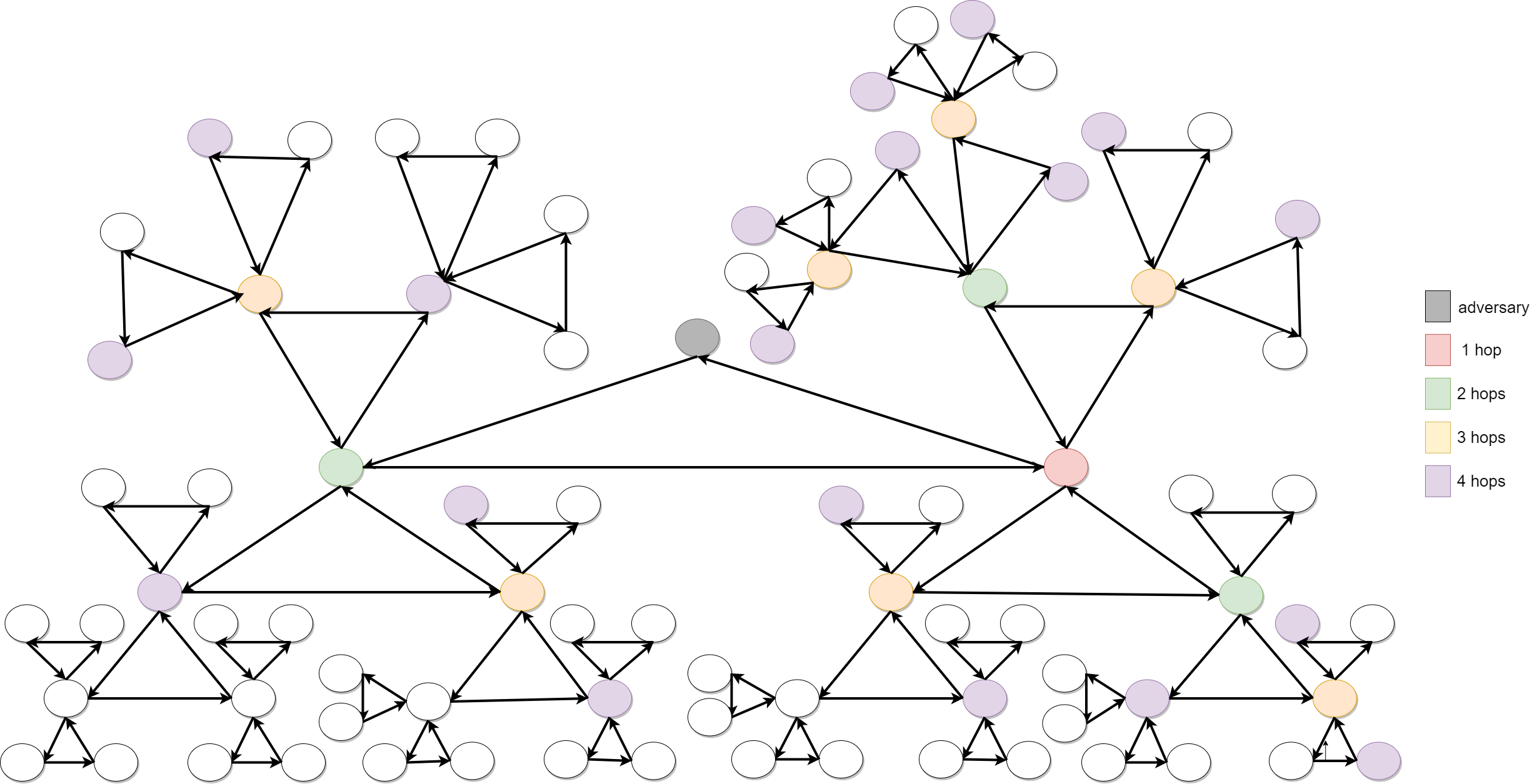}
    \caption{Extension of Figure \ref{fig:Uncertainty2hops_ctarzan_degree3}}
    \label{fig:CTarzanAS_3}
\end{figure*}

We start by considering the uncertainty at two hops in the standard Tarzan topology and a two-hop equivalent topology in which cycles are enabled.
This is represented in Figure \ref{fig:Uncertainty2hops}. Specifically, in Figure \ref{fig:Uncertainty2hops_tarzan_degree3}, we represent the standard Tarzan topology in which each node has three mimics.
Suppose that the grey node receives a message from the red node. In this case, the candidate senders, at a maximum distance of two hops, are the red node and the two green nodes.

The same uncertainty is obtained in the cyclic topology represented in Figure \ref{fig:Uncertainty2hops_ctarzan_degree2} in which, again, the candidate senders, at a maximum distance of two hops, are the red node and the two green nodes.

Regarding the cover traffic, we observe that in Figure \ref{fig:Uncertainty2hops_tarzan_degree3}, we have three bidirectional links while in Figure \ref{fig:Uncertainty2hops_ctarzan_degree2} we have four unidirectional links, thus saving two unidirectional links.
Therefore, it appears that keeping the same uncertainty, we have a significant reduction of the cover traffic.

Unfortunately, we can realize that the growth of the size of the anonymity set 
for the cyclic approach is slightly slower than that of standard Tarzan.
We can understand this just by considering the case of
tunnel length equal to four.
To see this, we extend the topologies of Figures \ref{fig:Uncertainty2hops_tarzan_degree3} and \ref{fig:Uncertainty2hops_ctarzan_degree2}, in Figures \ref{fig:TarzanAS} and \ref{fig:CTarzanAS_2} respectively, to include 
tunnels with a maximum length of four hops.

In this case, the anonymity set of Figure \ref{fig:TarzanAS} contains 15 nodes, while the anonymity set of Figure \ref{fig:CTarzanAS_2} contains 11 nodes.

Moreover, we have to take into account also the price in terms of latency required in the cyclic approach.
However, the advantage in terms of cover traffic is maintained with respect to Tarzan.
Therefore, it is interesting to understand what happens if we 
compare the standard Tarzan with the cyclic version by considering
two topologies that determine the same cover traffic.

The effect at two hops is highlighted in Figure  \ref{fig:Uncertainty2hops_ctarzan_degree3} in which there are 6 unidirectional links equivalent to three bidirectional links of Tarzan. Therein, we can see that the candidate senders are the red node and the three green nodes. Therefore, the uncertainty at two hops is increased.

The extension to four hops of Figure \ref{fig:Uncertainty2hops_ctarzan_degree3} is represented in Figure \ref{fig:CTarzanAS_3}.
In this case, the anonymity set contains 30 nodes. Therefore, under the same cover 
traffic, the cyclic approach offers a greater anonymity set size.
However, the price in terms of latency still remains.

Clearly, in Tarzan, the latency 
depends only on the tunnel length.
In the cyclic approach, it mostly depends on the tunnel length,
and in a small measure also depends on the node degree
(as explained in Section \ref{sec:latency}).
Moreover, the disadvantage of the cyclic version
depends also on the balance between downlink and uplink traffic
(the more the weight of the downlink, the more the disadvantage).

In fact, the price we pay in terms of latency is related to the downlink traffic for the return path, which is in general longer than the forward path.

Thus, the problem we want to study is the following: In the cyclic approach, can we reduce the tunnel length to reduce latency and still be able to have an anonymity set size greater than Tarzan?

If in general, the answer to this question could be negative, it is interesting to understand what happens when there is an unbalance between the quantity of uplink and downlink traffic. 

As we will describe in the sequel of the paper, the result we achieve is that for uplink-intensive networks, the above approach is definitely advantageous.

\section{C-Tarzan} \label{sec:C-Tarzan}
In this section, we propose a new protocol, called \textit{Circular Tarzan (C-Tarzan)}, based on the cyclic approach introduced in the previous section.

The idea is to move from from bidirectional links (adopted in Tarzan) to unidirectional links.

This is possible if the response is routed
through cycles which mimics belong to. As discussed above, we consider cycles of three nodes to minimize
the price in terms of latency.

To build the cycles among mimics nodes, we design a new mimic selection algorithm that differs from that of Tarzan.

We assume that the same Tarzan DHT table (with the $lookup$ function) is used in C-Tarzan for the mimic selection.


Each node $a$ chooses $k'$ mimics through the $lookup$ function (see Section \ref{sec:background}) as in Tarzan. Specifically, $a$ selects $b_i=lookup^i(a.ipaddr)$ for $1<i \leq k'+1$. Each chosen mimic $b_i$ can verify the correctness of the selection.
Anyway, differently from Tarzan, a unidirectional link directed from $a$ to $b_i$ is established.
At this point, each $b_i$ will chose a mimic $c_i=lookup^i(a.ipaddr||b_i.ipaddr)$ and a unidirectional link directed from $b_i$ to $c_i$ is established.
Observe that since the function $lookup$ accepts any arbitrary string as input and returns an IP address of a node of the network, it is guaranteed that the node $c_i$ always exists in the network. 
$c_i$ can verify the correctness of the mimic selection started by $a$, involving the node $b_i$. Finally, to close the cycle, a unidirectional link is established from $c_i$ to $a$. 

It is easy to realize that each node has on average $6k'$ mimics.
Indeed, each node $A$ \textit{selects directly} $k'$ mimics $B_1,\ldots B_{k'}$ to build $k'$ cycles. In each cycle involving the node $B_i$, there will be a node $C_i$ that establishes a link with $A$ to close the cycle. Then, $A$ will have further $k'$ mimics $C_1,\ldots C_{k'}$, resulting in a total of $2k'$ mimics.
At this point, on average, $A$ \textit{is selected directly} by $k'$ nodes to build further $k'$ cycles. This leads to further $2k'$ mimics for $A$. 
Finally, on average, $A$ \textit{is selected indirectly} by $k'$ nodes that in turn are selected directly by other $k'$ nodes to build cycles. As before, this results in further $2k'$ mimics for $A$.
Therefore, since unidirectional links are established between pairs of mimics, each node has, on average, $6k'$ unidirectional links ($3k'$ outgoing and $3k'$ ingoing).

We recall that, in Tarzan, if a node selects $k$ mimics, it has, on average, $2k$ mimics and then $2k$ bidirectional links corresponding to $4k$ unidirectional links. 
Therefore, by considering the number of links as a measure of cover traffic, we have that, to obtain the same level of cover traffic in Tarzan and C-Tarzan, we have to set $k'$ such that $6\cdot k'=4\cdot k$ i.e., $k'=\frac{2}{3} \cdot k$.

At this point, we discuss how the messages are forwarded anonymously towards the destination and the latter can reply to the initiator.

As in Tarzan, we assume that a symmetric hop-by-hop key is exchanged preliminarily between mimics. 

To enable the communication, we need to redefine the entire building process of the tunnel. Specifically, the initiator $a$ selects, as first relay, one of its \textit{outgoing} mimics $b_i$, i.e., a mimic $b_i$ such that a directed link from $a$ to $b_i$ exists.
Similarly to the standard Tarzan protocol, $a$ needs the set of the (outgoing) mimics of $b_i$ and to exchange a symmetric key with $b_i$. Anyway, since the link between $a$ and $b_i$ is unidirectional, a reply cannot be sent directly from $b_i$ to $a$, because it would be not covered by dummy traffic.

Therefore, to enable the reply, we define the function $C.next$ that can be invoked by a node $C$. This function receives as input a node $B$ and returns as output the node $A$, such that there exist: (i) a direct link from $B$ to $C$, (ii) a direct link from $C$ to $A$, (iii) a direct link from $A$ to $B$. 
Observe that, the $next$ function leverages on the fact that each node locally stores all the cycles it belongs to. Therefore, for a node $C$, given a node $B$ as input, it is straightforward to compute the next of the node $C$ (i.e., $A=C.next(B)$) in the cycle $BCAB$.  

Then, $b_i$ encrypts the response for $a$ by using the hop-by-hop key exchanged with $a$ and forwards this message to $c_i=b_i.next(a)$. This encrypted message is encrypted, in turn, by $b_i$ with the hop-by-hop key exchanged with $c_i$. 
At this point, $c_i$ decrypts the message, invokes the function $next$ to retrieve $a=c_i.next(b_i)$, encrypts the message again with its hop-by-hop key exchanged with $a$, and forwards it to $a$. 
Observe that, even though $c_i$ knows that some real traffic has to be forwarded to $a$ from $b_i$, $c_i$ does not know the content of it, and then it has no more information than $b_i$ about the fact that $a$ is the actual initiator or just an intermediate node of the tunnel.

Once obtained the outgoing mimics of $b_i$, $a$ selects a new mimic among them, say $d_i$, and needs to exchange a symmetric key and the set of outgoing mimics of $d_i$.
Now, two cases may occur. The first case is that $d_i=c_i$ i.e., $a, b_i, d_i$ are in the same cycle and $d_i$ coincides with $c_i$. 
In this case, the list of mimics of $c_i$ can be communicated directly through the link between $c_i$ and $a$.

The second (complementary) case occurs when $d_i$ has no common cycle with $a$. In this case, the list of mimics has to be forwarded from $d_i$ to $a_i$ through $b_i$. 
To enable the communication between $d_i$ and $b_i$, since no direct link exists from $d_i$ to $b_i$, we apply the approach discussed above. Specifically, $d_i$ forwards this list through another node $e_i=d_i.next(b_i)$.

\begin{figure}
     \centering
     \begin{subfigure}[b]{0.4\textwidth}
         \centering
         \includegraphics[width=0.6\textwidth]{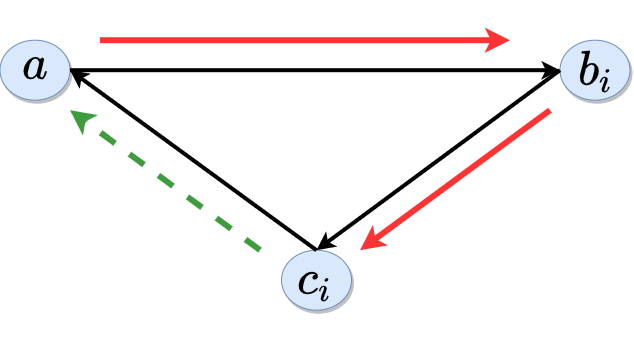}
         \caption{Second relay in the same cycle of the initiator}
         \label{fig:responsea}
     \end{subfigure}
     \begin{subfigure}[b]{0.5\textwidth}
         \centering
         \includegraphics[width=\textwidth]{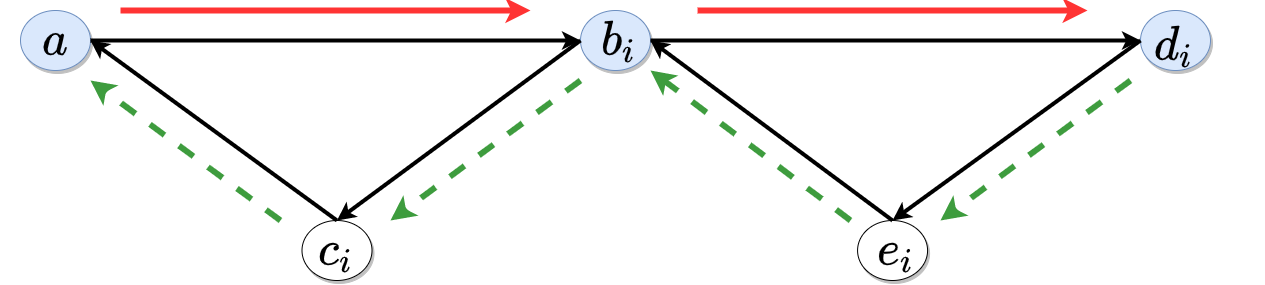}
         \caption{Second relay in a different cycle from the initiator}
         \label{fig:responseb}
     \end{subfigure}
        \caption{Second relay selection}
        \label{fig:response}
\end{figure}

These two cases are represented in Figures \ref{fig:responsea} and \ref{fig:responseb}, respectively. Therein, we represent by a red arrow the forward communication between the initiator and the second relay of the tunnel, and by a green dashed arrow the backward communication from the second relay to the initiator.
 
The building of the tunnel proceeds iteratively until the last node.

Once the tunnel is set, the initiator can communicate with the recipient through this tunnel as in the standard Tarzan protocol. 

Regarding the response by the recipient, the approach used to enable the exchange of information between a node of the tunnel and a previous node is applied.
Specifically, at each hop of the tunnel starting from the last node until the initiator, if a direct link exists between a node and a previous node of the tunnel, then the response is directly forwarded through this link, otherwise the response is forwarded through an intermediate node.

Some more detail will be discussed in Section \ref{sec:latency}.

\section{Latency in Tarzan and C-Tarzan} \label{sec:latency} 

In the previous sections, we mentioned that our solution introduces a price in terms of latency, assuming the same cover traffic and the same tunnel length in Tarzan and C-Tarzan. To give an answer to the question of Section \ref{sec:problem_formulation}, we have to quantify this price.

To perform an analytic analysis, we use as a measure of this metric the number of hops travelled
by a message in the forward path and in the return path.

We introduce the following notation. We denote by $\tau$ the average delay of the links of the network.
We start by evaluating the latency for Tarzan.
We denote by $h$ the tunnel length of Tarzan and by $L_f$ and $L_r$ the latency of the forward path and the latency of the return path of Tarzan, respectively.

Since the same tunnel is used both for the request and the response, it is easy to see that $L_f=L_r=(h+2)\cdot \tau$, where the term $2$ derives from the fact that there is one hop between the last node of the tunnel and the PNAT and one hop from the PNAT and the destination.

Consider now C-Tarzan. We denote by $h'$ the tunnel length and by $L'_f$ and $L'_r$ the latency of the forward path and the latency of the return path, respectively.

For the forward path, no difference with Tarzan exists and then $L'_f=(h'+2)\cdot \tau$.

On the other hand, for the return path, it is not trivial to estimate the number of hops in the return path, since it depends on the tunnel construction.

We can provide an approximation of the return latency representing an upper bound of its actual value.

The two cases of Figures \ref{fig:responsea} and \ref{fig:responseb} have to be considered.
In particular, consider the selection of the first two relays of the tunnel.
If the second relay is in the same cycle as the initiator (case a), then the response goes directly from the second relay $c_i$ to the initiator $a$ and this means that two hops in the forward path correspond to just one hop in the return path.

In case (b), the second relay $d_i$ belongs to a different cycle and then the response goes from $d_i$ to the first relay $b_i$, through an intermediate node $e_i$ (2 hops) and, then, from $b_i$ to the initiator $a$, through another intermediate node $c_i$ (again, 2 hops).

In other words, for the first two hops of the forward phase, if case (a) occurs, then the response requires one hop, otherwise (case (b)), the response requires four hops. 

It remains to estimate the probability that cases (a) and (b) occur.

To do this, we denote by $d=3k'$ the average number of outgoing mimics of a node.

Obviously, since the mimics are selected uniformly at random, the case (a) occurs with probability $\frac{1}{d}$ and the case (b) occurs with probability $\frac{d-1}{d}$.

So far, no approximation has been introduced. 

If we assume that the third relay of the tunnel is selected in a different cycle than the first relay (it happens with probability $\frac{d-1}{d}$), we can apply the reasoning followed for the first two relays to the third and fourth relays.
Therefore, to find an approximation, we neglect the event that the relays in an odd position $i$ of the tunnel are selected in the same cycle of the relay in position $i-2$ (it happens with probability $\frac{1}{d}$ at each choice).

Under this hypothesis, we have that, for every two hops in the forward phase, if case (a) occurs, then the response requires one hop, otherwise (case (b)), the response requires four hops. 

It is easy to realize that, if such a hypothesis is not satisfied, then the response requires a lower number of hops and then, our approximation represents an upper bound of the actual latency.

Therefore, for $h'$ even, the latency of the return path of C-Tarzan results: $(\frac{h'}{2} \cdot (\frac{1}{d} \cdot 1 + \frac{d-1}{d} \cdot 4)+2)   \cdot \tau= (h' \cdot (2-\frac{3}{2 \cdot d})+2) \cdot \tau$. 
 
On the other hand, for $h'$ odd, the latency results: $( (h'-1) \cdot (2-\frac{3}{2 \cdot d})+4) \cdot \tau$.

By considering equally likely the events that $h'$ is odd and $h'$ is even, we conclude that the return latency for C-Tarzan is:
$L'_r=(h' \cdot (2-\frac{3}{2\cdot d})+ \frac{3}{4 \cdot d}+2) \cdot \tau$.


Observe that $L'_r$ increases as $d$ increases. This is due to the fact that, as $d$ increases, the probability that a mimic of the tunnel is selected in a different cycle increases. Then, the response requires more hops and the return latency increases.


\section{Experiments} \label{sec:experiments}
Through this section, we perform an experimental validation of C-Tarzan by highlighting the conditions under which it outperforms the standard Tarzan protocol.

\subsection{Metrics and Experiment Setting}
As already introduced, we consider three metrics: cover traffic, latency, and anonymity set size.

Regarding the cover traffic, we use as a measure the number of ingoing and outgoing links of the nodes,
by considering that every link concurs, in the average, with the same portion of cover traffic.
As discussed in Section \ref{sec:C-Tarzan}, to obtain the same cover traffic in Tarzan and C-Tarzan, we have to set 
\begin{equation} \label{eq:kCTarzan}
k'=\frac{2}{3}\cdot k
\end{equation}
Regarding the latency, as seen in Section \ref{sec:latency}, to obtain the same total latency (forward latency plus return latency) we need to set $h'$ such that $L_f+L_r=L'_f+L'_r$ i.e., $h'= \frac{2h-\frac{3}{4\cdot d}}{3- \frac{3}{2\cdot d}}$.
However, since we are interested in studying what happens when the balance
between uplink and downlink traffic varies, we introduce two coefficients $w_f$ and $w_r$, such that $w_f+w_r=2$, to associate with the forward latency and the return latency, respectively.
For example, $w_f=w_r=1$ represents a balanced traffic between uplink and downlink, while $w_f=2$ and $w_r=0$ represents only uplink traffic.

Therefore, the condition to satisfy is $w_f \cdot L'_f+ w_r \cdot L'_r=w_f \cdot L_f+ w_r \cdot L_r$, that leads to
\begin{equation} \label{eq:hCTarzan}
h'=\frac{2 \cdot h-\frac{3}{4 \cdot d} \cdot w_r}{w_f+ \frac{4 \cdot d-3}{2 \cdot d}\cdot w_r}
\end{equation}

Now, we denote by $AS(k,h)$ the size of the anonymity set of Tarzan obtained as a function of $k$ and $h$.
Furthermore, we denote by $AS'(k',h')$ the size of the anonymity set of C-Tarzan obtained as a function of $k'$ and $h'$.

Thus, the question now is whether, by setting $k'$ and $h'$ as in equations \ref{eq:kCTarzan} and \ref{eq:hCTarzan}, respectively, it holds that $AS'$ is greater than $AS$.
If this is the case, then our approach introduces an advantage with respect to Tarzan.

Due to the complexity of retrieving the analytical formulas for $AS$ and $AS'$, we do this by simulation,
leaving the analytical study as a future work.

Furthermore, in order to obtain realistic results, we do not use directly the upper bound provided by \ref{eq:hCTarzan} (see Section \ref{sec:latency}), but we find experimentally the values of $h$ and $h'$ leading to the same latency for Tarzan and C-Tarzan, respectively
(actually, verifying the results obtained in Section \ref{sec:latency}).

To summarize, we find the values $(h,k,h',k')$ that satisfy the following system.


     

\begin{equation} \label{eq:system}
    \begin{cases}
     k'=\frac{2}{3} \cdot k\\
     w_f+w_r=2 \\
     w_f \cdot L'_f+ w_r \cdot L'_r=w_f \cdot L_f+ w_r \cdot L_r\\
     AS' \geq AS
     
    \end{cases}
\end{equation}

In detail, the simulation has been performed in JAVA as follows. We considered a network of $100,000$ nodes.
First, we set some values of $w_f$ (and, then, $w_r=2-w_f$), $k'$, and $h'$ for C-Tarzan and,
then, we generated a topology (the links are obtained considering that each node selects \textit{directly} $k'$ mimics to build cycles).

On this topology, we measured the average degree of each node counting both the actual ingoing and the outgoing links (cover traffic), the actual number of hops that a request and the corresponding response have to cross on a path of height $h'$ (measure of latency), and the size of the corresponding anonymity set.

We repeated the experiment with the same parameters for 100 rounds (by varying the topology) to obtain steady results.


At this point, by the first equation of the system (\ref{eq:system}), we set $k=\frac{3}{2} \cdot k'$.
Then, by using the value $w_f \cdot L'_f+ w_r \cdot L'_r$ obtained experimentally for C-Tarzan and by recalling that $L_f=L_r=(h+2) \cdot \tau$, by the second and third equations of the system (\ref{eq:system}), we found the proper value of $h=\frac{w_f \cdot L'_f+ w_r \cdot L'_r-4 \cdot \tau}{2 \cdot \tau}$.

Then, we performed again 100 rounds of simulation with $k, h$ to measure the cover traffic, latency, and anonymity set of Tarzan.

We confirmed that the obtained values of cover traffic and latency are the same as C-Tarzan (with an error less than 1 \% for both). Therefore, we obtain
an experimental validation of the fact that the first three equations of ($\ref{eq:system}$) hold.
We discuss the results regarding the anonymity set size in the next section.

\subsection{Results} \label{sec:results}

In this section, we compare Tarzan and C-Tarzan in terms of anonymity set size, by setting the same cover traffic and same latency.

In the first analysis, we show as the anonymity set size of both the protocols varies as the cover traffic increases.
We plot in the $y$-axis the ratio between the size of the anonymity set of C-Tarzan $AS'$ and the size of the anonymity set of Tarzan $AS$. In the $x$-axis, we consider the degree $d$ representing the number of outgoing (or ingoing) links in C-Tarzan (as defined in Section \ref{sec:latency}) that is equal to the number of bidirectional links in Tarzan (to obtain the same cover traffic).

The results of this analysis are reported in Figures \ref{fig:anonymitySetVSd_h3},\ref{fig:anonymitySetVSd_h4},\ref{fig:anonymitySetVSd_h5}, for different values of $h'$ and $w_f$.

\begin{figure}%
\centering%
\begin{gnuplot}[terminal=epslatex, terminaloptions=color dashed]

set termoption dash

set style line 1 lt rgb "black" lw 5 pt 3  dt 2
#set style line 2 lt rgb "web-green" lw 5 pt 3 
#set style line 3 lt rgb "blue" lw 9 pt 2 dt 2
#set style line 4 lt rgb "blue" lw 5 pt 2
#set style line 5 lt rgb "web-green" lw 3 pt 3  dt 2
#set style line 6 lt rgb "web-green" lw 5 pt 3

set size 0.720,0.800
#set xtics  12,5,56
set key inside top right
set xlabel "d" 
set ylabel "AS'/AS" off 1
#set title "h=3"

plot  "h3.txt" using 1:2 title 'Threshold' with linespoints ls 1, "h3.txt" using 1:3 title 'wf=1.5' with linespoints, "h3.txt" using 1:4 title 'wf=1.6' with linespoints, "h3.txt" using 1:5 title 'wf=1.7' with linespoints, "h3.txt" using 1:6 title 'wf=1.8' with linespoints, "h3.txt" using 1:7 title 'wf=1.9' with linespoints
\end{gnuplot}
\caption{Anonymity set ratio vs cover traffic $d$ with $h'$=3}%
\label{fig:anonymitySetVSd_h3}%
\end{figure}%

\begin{figure}%
\centering%
\begin{gnuplot}[terminal=epslatex, terminaloptions=color dashed]

set termoption dash

set style line 1 lt rgb "black" lw 5 pt 3  dt 2
#set style line 2 lt rgb "web-green" lw 5 pt 3 
#set style line 3 lt rgb "blue" lw 9 pt 2 dt 2
#set style line 4 lt rgb "blue" lw 5 pt 2
#set style line 5 lt rgb "web-green" lw 3 pt 3  dt 2
#set style line 6 lt rgb "web-green" lw 5 pt 3

set size 0.720,0.800
#set xtics  12,5,56
set key inside top right
set xlabel "d" 
set ylabel "AS'/AS" off 1
#set title "h=3"

plot  "h4.txt" using 1:2 title 'Threshold' with linespoints ls 1, "h4.txt" using 1:3 title 'wf=1.5' with linespoints, "h4.txt" using 1:4 title 'wf=1.6' with linespoints, "h4.txt" using 1:5 title 'wf=1.7' with linespoints, "h4.txt" using 1:6 title 'wf=1.8' with linespoints, "h4.txt" using 1:7 title 'wf=1.9' with linespoints
\end{gnuplot}
\caption{Anonymity set ratio vs cover traffic $d$ with $h'$=4}%
\label{fig:anonymitySetVSd_h4}%
\end{figure}%

\begin{figure}%
\centering%
\begin{gnuplot}[terminal=epslatex, terminaloptions=color dashed]

set termoption dash

set style line 1 lt rgb "black" lw 5 pt 3  dt 2
#set style line 2 lt rgb "web-green" lw 5 pt 3 
#set style line 3 lt rgb "blue" lw 9 pt 2 dt 2
#set style line 4 lt rgb "blue" lw 5 pt 2
#set style line 5 lt rgb "web-green" lw 3 pt 3  dt 2
#set style line 6 lt rgb "web-green" lw 5 pt 3

set size 0.720,0.800
#set xtics  12,5,56
set key inside top right
set xlabel "d" 
set ylabel "AS'/AS" off 1
#set title "h=3"

plot  "h5.txt" using 1:2 title 'Threshold' with linespoints ls 1, "h5.txt" using 1:3 title 'wf=1.5' with linespoints, "h5.txt" using 1:4 title 'wf=1.6' with linespoints, "h5.txt" using 1:5 title 'wf=1.7' with linespoints, "h5.txt" using 1:6 title 'wf=1.8' with linespoints, "h5.txt" using 1:7 title 'wf=1.9' with linespoints
\end{gnuplot}
\caption{Anonymity set ratio vs cover traffic $d$ with $h'$=5}%
\label{fig:anonymitySetVSd_h5}%
\end{figure}%

We represent with a dashed black line the ratio equal to $1$. When the plots exceed this line, C-Tarzan outperforms Tarzan (in terms of anonymity set size).

We observe that our performance (for a fixed $h'$) decreases as $d$ increases.

This happens because, as $d$ increases, the latency of C-Tarzan increases, then the tunnel length of Tarzan $h$ (that offers the same latency of C-Tarzan) increases too. Therefore, the anonymity set size of Tarzan increases.

Even though the anonymity set size of both the protocols has a polynomial growth with $d$, the exponential growth of the anonymity set size of Tarzan with $h$ is dominant. Therefore, as $d$ increases, the ratio between $AS'$ and $AS$ decreases.

Regarding $w_f$, as it increases (by considering the same $d$), the performance of C-Tarzan increases.
This happens because an increasing weight $w_f$ represents predominant uplink traffic that leads to lower total latency for C-Tarzan (since the return path is longer than the forward path). This implies that the tunnel length $h$ of Tarzan, which offers the same latency, decreases and then $AS$ decreases too.

As a final consideration, we observe that, until a certain level of cover traffic (corresponding to some $d$), it is advantageous to employ the C-Tarzan protocol, while when this threshold is exceeded, Tarzan is more convenient.  
Moreover, in the condition of increasing uplink traffic, this threshold also increases by making C-Tarzan suitable within a higher range of cover traffic level.

Observe that lower values of $d$ are desirable since they represent cover traffic injected in the network. On the other hand, the reader might ask whether lower values of $d$ results in acceptable anonymity set size in absolute terms (in relative terms C-Tarzan outperforms Tarzan). The response is affirmative, indeed as we discuss in the sequel, the anonymity set increases exponentially with $h$ and $h'$. Then, with a small increment of $h'$, we are able to obtain good anonymity set size still outperforming Tarzan. Just an example, with $d=4$ and $h'=4$, we obtain an anonymity set size of about $100$.   

We conclude this section, by showing as the performances of C-Tarzan vary with respect to Tarzan as $h'$ varies.  

The plot in Figure \ref{fig:anonymitySet_d4} shows $AS$ and $AS'$ as $h'$ varies with two different values of $w_f$ and $d=4$.

As expected, $AS'$ increases exponentially with $h'$. 
Moreover, when $h'$ increases, $h$ increases too (to offer the same latency), and then also $AS$ increases exponentially.

Observe that $AS'$ with $w_f=1.5$ is essentially (modulo experimental error) the same as $AS'$ with $w_f=1.9$. Indeed, $AS'$ does not depend on $w_f$.

On the contrary, $h$ depends on the total latency of Tarzan, which is equal to the total latency of C-Tarzan that, in turn, depends on $w_f$. Therefore, as $w_f$ increases, $h$ decreases and $AS$ decreases too.

To conclude this section, in Figures \ref{fig:anonymitySetVSh_d3}, \ref{fig:anonymitySetVSh_d4}, and \ref{fig:anonymitySetVSh_d5},  we show the ratio between the anonymity set of Tarzan and C-Tarzan as $h'$ varies for different values of $w_f$ and $d$. 

\begin{figure}%
\centering%
\begin{gnuplot}[terminal=epslatex, terminaloptions=color dashed]

set termoption dash

set style line 1 lt rgb "red" lw 5 pt 3  dt 2
set style line 2 lt rgb "blue" lw 5 pt 4 dt 2
set style line 3 lt rgb "red" lw 5 pt 2 
set style line 4 lt rgb "blue" lw 5 pt 5
#set style line 5 lt rgb "web-green" lw 3 pt 3  dt 2
#set style line 6 lt rgb "web-green" lw 5 pt 3

set size 0.720,0.800
#set xtics  12,5,56
set key inside top left
set xlabel "h'"
set ylabel "Anonymity Set" off 1
#set title "d=3"

plot  "anonimity.txt" using 1:10 title 'Tarzan wf=1.9' with linespoints, "anonimity.txt" using 1:2 title 'Tarzan wf=1.5' with linespoints, "anonimity.txt" using 1:3 title 'C-Tarzan wf=1.5' with linespoints, "anonimity.txt" using 1:11 title 'C-Tarzan wf=1.9' with linespoints
\end{gnuplot}
\caption{Anonymity set vs $h'$ with $d$=4}%
\label{fig:anonymitySet_d4}%
\end{figure}%

\begin{figure}%
\centering%
\begin{gnuplot}[terminal=epslatex, terminaloptions=color dashed]

set termoption dash

set style line 1 lt rgb "black" lw 5 pt 3  dt 2
#set style line 2 lt rgb "web-green" lw 5 pt 3 
#set style line 3 lt rgb "blue" lw 9 pt 2 dt 2
#set style line 4 lt rgb "blue" lw 5 pt 2
#set style line 5 lt rgb "web-green" lw 3 pt 3  dt 2
#set style line 6 lt rgb "web-green" lw 5 pt 3

set size 0.720,0.800
#set xtics  12,5,56
set key inside top left
set xlabel "h'"
set ylabel "AS'/AS" off 1
#set title "d=3"

plot  "d3.txt" using 1:2 title 'Threshold' with linespoints ls 1, "d3.txt" using 1:3 title 'wf=1.5' with linespoints, "d3.txt" using 1:4 title 'wf=1.6' with linespoints, "d3.txt" using 1:5 title 'wf=1.7' with linespoints, "d3.txt" using 1:6 title 'wf=1.8' with linespoints, "d3.txt" using 1:7 title 'wf=1.9' with linespoints
\end{gnuplot}
\caption{Anonymity set ratio vs tunnel length $h'$ with $d$=3}%
\label{fig:anonymitySetVSh_d3}%
\end{figure}%

\begin{figure}%
\centering%
\begin{gnuplot}[terminal=epslatex, terminaloptions=color dashed]

set termoption dash

set style line 1 lt rgb "black" lw 5 pt 3  dt 2
#set style line 2 lt rgb "web-green" lw 5 pt 3 
#set style line 3 lt rgb "blue" lw 9 pt 2 dt 2
#set style line 4 lt rgb "blue" lw 5 pt 2
#set style line 5 lt rgb "web-green" lw 3 pt 3  dt 2
#set style line 6 lt rgb "web-green" lw 5 pt 3

set size 0.720,0.800
#set xtics  12,5,56
set key inside top left
set xlabel "h'"
set ylabel "AS'/AS" off 1
#set title "d=4"

plot  "d4.txt" using 1:2 title 'Threshold' with linespoints ls 1, "d4.txt" using 1:3 title 'wf=1.5' with linespoints, "d4.txt" using 1:4 title 'wf=1.6' with linespoints, "d4.txt" using 1:5 title 'wf=1.7' with linespoints, "d4.txt" using 1:6 title 'wf=1.8' with linespoints, "d4.txt" using 1:7 title 'wf=1.9' with linespoints
\end{gnuplot}
\caption{Anonymity set ratio vs tunnel length $h'$ with $d$=4}%
\label{fig:anonymitySetVSh_d4}%
\end{figure}%

\begin{figure}%
\centering%
\begin{gnuplot}[terminal=epslatex, terminaloptions=color dashed]

set termoption dash

set style line 1 lt rgb "black" lw 5 pt 3  dt 2
#set style line 2 lt rgb "web-green" lw 5 pt 3 
#set style line 3 lt rgb "blue" lw 9 pt 2 dt 2
#set style line 4 lt rgb "blue" lw 5 pt 2
#set style line 5 lt rgb "web-green" lw 3 pt 3  dt 2
#set style line 6 lt rgb "web-green" lw 5 pt 3

set size 0.720,0.800
#set xtics  12,5,56
set key inside top left
set xlabel "h'"
set ylabel "AS'/AS" off 1
#set title "d=5"

plot  "d5.txt" using 1:2 title 'Threshold' with linespoints ls 1, "d5.txt" using 1:3 title 'wf=1.5' with linespoints, "d5.txt" using 1:4 title 'wf=1.6' with linespoints, "d5.txt" using 1:5 title 'wf=1.7' with linespoints, "d5.txt" using 1:6 title 'wf=1.8' with linespoints, "d5.txt" using 1:7 title 'wf=1.9' with linespoints
\end{gnuplot}
\caption{Anonymity set ratio vs tunnel length $h'$ with $d$=5}%
\label{fig:anonymitySetVSh_d5}%
\end{figure}%

According to the previous analysis, C-Tarzan outperforms Tarzan for low $d$ and for increasing $w_f$.
Regarding $h'$, we observe a fluctuating behaviour in which there are some ranges of $h$ in which there is an increasing trend of the ratio and other ranges in which there is an opposite trend.
This is due to a compensation effect between the growth of the anonymity set size and the latency.
In particular, for C-Tarzan, when $h'$ increases, $AS'$ increases, and the total latency increases too. Anyway, in some ranges, the increment of latency is limited. This leads to an increment of the tunnel length of Tarzan $h$ that is not sufficient to obtain an anonymity set size $AS$ which compensates for the growth of $AS'$.

On the contrary, once $h'$ reaches a peak value, the effect of the growth of the latency assumes a more relevant role by leading to values of $h$ corresponding to anonymity set size $AS$ able to compensate for the growth of $AS'$.

As a final remark, observe that, in this analysis, we show the advantage of our approach just in terms of anonymity set size (under the same latency and cover traffic level). Clearly, this advantage can be translated into an advantage in terms of latency or cover traffic, by fixing the same anonymity set size for both the protocols.

\section{Conclusion} \label{sec:conclusion}

In this paper, we presented C-Tarzan, a new P2P overlay anonymous protocol Tarzan for uplink-intensive applications.

The performance of C-Tarzan is evaluated according to the three main metrics \cite{das2018anonymity} in the field of anonymous communications: latency, amount of cover traffic, and anonymity set size.
We performed an in-depth experimental validation highlighting the conditions under which it is more advantageous to employ C-Tarzan instead of Tarzan.
The main result, arising from the conducted analysis, is that C-Tarzan outperforms Tarzan in terms of anonymity set size
above a certain uplink-traffic threshold, until a 
relevant improvement for uplink-intensive applications.
A direction to investigate as a future work consists of extending the experimental validation by including a network simulation that takes into account some more detail (such as the queuing of the packets in the node buffers). Anyway, we expect that the above comparative results are not affected by the above factors because of the substantial similarity between the protocols Tarzan and C-Tarzan concerning the network aspects.
In fact, we are performing a simulation through NS3, and from the first very preliminary results, that we did not include for space limitation, it appears that the results of Section \ref{sec:experiments} are fully confirmed.

\end{document}